\documentclass[12pt,a4paper,DIV13]{scrartcl}

\usepackage[UKenglish]{babel}
\usepackage[latin1]{inputenc}
\usepackage{graphicx} 
\usepackage{scrpage2}
\usepackage{indentfirst}
\usepackage{amsmath} 
\usepackage{booktabs} 
\usepackage{array} 
\usepackage{upgreek} 
\usepackage[font=small]{caption}
\usepackage{subfigure}
\usepackage{cite} 
\usepackage[figurename=FIG.]{caption} 
\usepackage[tablename=TABLE]{caption} 
\usepackage[version=3]{mhchem} 
\usepackage{textcomp} 
\usepackage{footnote}
\makesavenoteenv{tabular}
\DeclareOldFontCommand{\sf}{\normalfont\sffamily}{\mathsf}
\DeclareOldFontCommand{\bf}{\normalfont\bfseries}{\mathbf}
\DeclareOldFontCommand{\sfb}{\normalfont\bfseries\sffamily}{\mathsfb}
\DeclareOldFontCommand{\it}{\normalfont\itshape}{\mathit}

\renewcommand{\thetable}{\Roman{table}} 
\renewcommand{\thesection}{\Roman{section}}
\renewcommand{\thesubsection}{\Alph{subsection}}
\renewcommand{\thefootnote}{\alph{footnote}}

\newcolumntype{C}[1]{>{\centering\arraybackslash}m{#1}}
\newcolumntype{L}[1]{>{\raggedright\let\newline\\\arraybackslash\hspace{0pt}}m{#1}}
\begin{document}
	
	\thispagestyle{empty}
	
	\begin{center}
		{\LARGE \sfb Electron effective mass and mobility limits in degenerate perovskite stannate BaSnO$_3$}
	\end{center}
	
	\vspace{0.5cm}
	
	\begin{flushleft}
		
		\begingroup\linespread{1.0}\selectfont
		Christian A. Niedermeier$^{1,2,3}\footnote{Corresponding author, e-mail: c.niedermeier13@imperial.ac.uk}$, Sneha Rhode$^{1}$, Keisuke Ide$^{2}$, Hidenori Hiramatsu$^{2,3}$, Hideo Hosono$^{2,3}$, Toshio Kamiya$^{2,3}$, Michelle A. Moram$^{1}$
		
		$^1$Department of Materials, Imperial College London, Exhibition Road, London, SW7~2AZ, UK
		
		$^2$Laboratory for Materials and Structures, Tokyo Institute of Technology, Mailbox R3-4, 4259 Nagatsuta, Midori-ku, Yokohama, 226-8503, Japan
		
		$^3$Materials Research Center for Element Strategy, Tokyo Institute of Technology, 4259 Nagatsuta, Midori-ku, Yokohama, 226-8503, Japan
		
		
		\endgroup
		
	\end{flushleft}
	
	\vspace{0cm}
	
	{\sfb Abstract:}
	The high room temperature mobility and the electron effective mass in BaSnO$_3$ are investigated in depth by evaluation of the free carrier absorption observed in infrared spectra for epitaxial films with free electron concentrations from $8.3 \times 10^{18}$ to $7.3 \times 10^{20}$~cm$^{-3}$. Both the optical band gap widening by conduction band filling and the carrier scattering mechanisms in the low and high doping regimes are consistently described employing parameters solely based on the intrinsic physical properties of BaSnO$_3$. The results explain the current mobility limits in epitaxial films and demonstrate the potential of BaSnO$_3$ to outperform established wide band gap semiconductors also in the moderate doping regime.
	
	\newpage
	\setcounter{page}{1}
	\setlength{\parindent}{10pt}
	
	Transparent perovskite stannate BaSnO$_3$ shows great potential as a high-mobility electron transport material composed of abundant elements. Since the report of an extraordinary high room temperature mobility of 320~cm$^2$/Vs for La:BaSnO$_3$ single crystals~\cite{Kim2012_APE}, which is the highest value reported for perovskite oxides, the material has rapidly attracted interest as high-mobility channel layer in oxide thin film transistors~\cite{Park2014,Kim2015,Kim2016} and multi-functional perovskite-based optoelectronic devices~\cite{Ismail-Beigi2015,Krishnaswamy2016}. To fully exploit the potential of La:BaSnO$_3$ for device applications, current research concentrates on understanding and improving the electron transport in epitaxial La:BaSnO$_3$ thin films~\cite{Kim2012_PRB, Mun2013, Sallis2013, Wadekar2014, Kim2014, Raghavan2016, Lee2016, Niedermeier2016}.
	
	La:BaSnO$_3$ thin films grown heteroepitaxially on SrTiO$_3$ substrates using pulsed laser deposition (PLD) show a reduced carrier mobility of 70~cm$^2$/Vs~\cite{Kim2012_PRB} as compared to single crystals. It was suggested that the high density of dislocations introduced scattering centres for electron transport~\cite{Mun2013}. However, even by homoepitaxy on BaSnO$_3$ single crystals the carrier mobility can be increased only to 100~cm$^2$/Vs~\cite{Lee2016}. Moreover, molecular beam epitaxy allows for preparation of La:BaSnO$_3$ films with the record Hall mobility of 124~cm$^2$/Vs on SrTiO$_3$ substrates~\cite{Raghavan2016}, suggesting the larger impact of the deposition method on electron transport properties. A quantitative analysis of the prevailing carrier scattering mechanism in BaSnO$_3$ is still lacking and would provide a guideline for improving the electron transport beyond the current mobility limits of epitaxial films.
	
	The high mobility in La:BaSnO$_3$ single crystals is attributed to the large dispersion of the Sn~5s~orbital-derived conduction band and the ideal 180$^\circ$ O$-$Sn$-$O bond angle in the network of corner sharing (SnO$_6)^{2-}$ octahedra in the cubic perovskite structure~\cite{Mizoguchi2004}. Quantitatively, the electron mobility is given by 
	\begin{equation}
	\mu = \frac{e \tau}{m_\text{e}^*}
	\label{eq:mobility}
	\end{equation}
	where $e$ is the electron charge, $m_\text{e}^*$ is the electron effective mass and $\tau$ is the relaxation time denoting the average time of momentum loss by scattering. A fundamental understanding of the electron transport in epitaxial La:BaSnO$_3$ films thus requires a quantitative analysis of the effective mass $m_\text{e}^*$ and scattering relaxation time $\tau$, which both strongly dependent on carrier concentration.\\
	
	This work presents an in-depth quantitative analysis of the La:BaSnO$_3$ electron effective mass by evaluation of the free carrier absorption observed in infrared spectra for a wide range of carrier concentrations from $8.3 \times 10^{18}$ to $7.3 \times 10^{20}$~cm$^{-3}$. The non-parabolicity of the La:BaSnO$_3$ conduction band is derived from the dependence of electron effective mass on doping level. The results are employed in an analytical model to consistently describe the optical band gap widening due to conduction band filling, by taking the band gap narrowing induced by electron-electron and electron-impurity interactions into account. The current mobility limits in La:BaSnO$_3$ epitaxial films are described well by the analytical models for electron scattering by longitudinal optical (LO) phonons, dislocations and ionized impurities for degenerate doping.
	
\begin{figure}[hp]
		\centering
		\includegraphics[width=\textwidth]{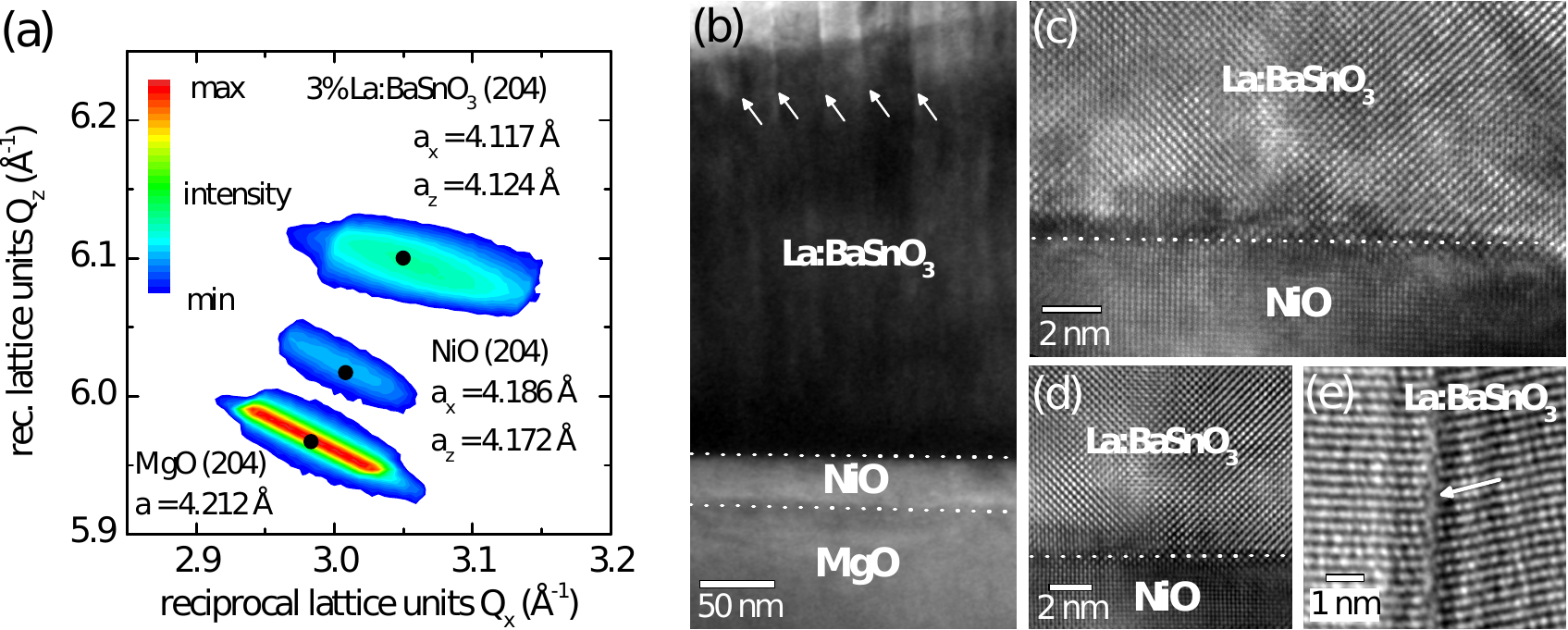}
		\caption{(a) HR-XRD 204 reciprocal space map of La:BaSnO$_3$/NiO/MgO given by an intensity contour map on a logarithmic scale with reference positions of unstrained crystals indicated by black dots. (b) Cross-sectional bright-field TEM micrograph indicates the presence of vertical grain boundaries in La:BaSnO$_3$ and columnar growth. (c)~Cross-sectional HR-TEM micrograph shows the single-domain epitaxy at the La:BaSnO$_3$/NiO interface which is free of misfit dislocations as confirmed by the (d) ABSF-filtered HR-TEM micrograph. (e) HR-TEM micrograph of a low-angle grain boundary between two La:BaSnO$_3$ crystallites.}
		\label{fig:tem}
	\end{figure}

	
	200-nm epitaxial La:BaSnO$_3$ thin films were grown by PLD on 50 nm NiO-buffered MgO substrates to reduce the lattice mismatch to less than 1.4\%. The experimental details for thin film growth are provided in the supplementary information~\cite{supplement}. The 204 reciprocal space map high resolution X-ray diffraction (HR-XRD) analysis confirms the 100 epitaxy of La:BaSnO$_3$ thin film on NiO-buffered MgO substrates~(Fig.~\ref{fig:tem}(a)). The NiO buffer layer is slightly strained in plane while the La:BaSnO$_3$ thin film is completely relaxed and shows a moderate degree of mosaicity as indicated by the broadened diffraction peak relative to that of the MgO single crystal.
	
	The cross-sectional bright field transmission electron microscopy (TEM) image of the La:BaSnO$_3$ thin film on a NiO-buffered MgO substrate indicates columnar growth as deduced from the observation of grain boundaries marked by the arrows in Fig.~\ref{fig:tem}(b). The cross-sectional high resolution micrograph shows that the La:BaSnO$_3$/NiO interface is free of misfit dislocations~(Fig.~\ref{fig:tem}(c)) as confirmed by the average background substraction filtered (ABSF) HR-TEM micrograph~(Fig.~\ref{fig:tem}(d)). Consistent with the mosaicity derived from HR-XRD analysis, the La:BaSnO$_3$ microstructure shows grains of ca. 30~nm size which are slightly tilted with respect to the [100] epitaxial orientation and forms low-angle grain boundaries (Fig.~\ref{fig:tem}(e)).
	
\begin{figure}[h]
		\centering
		\includegraphics[width=8.5cm]{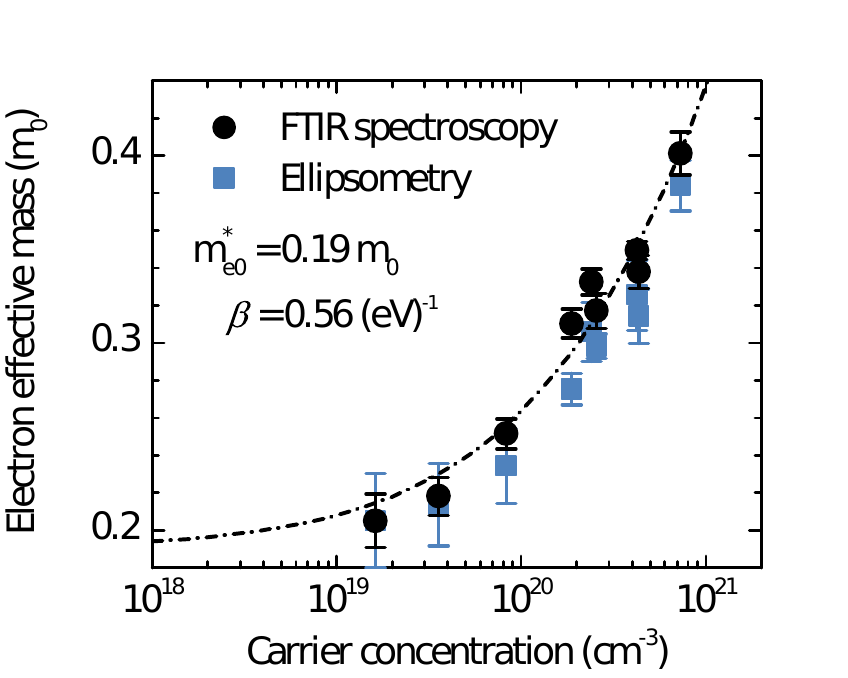}
		\caption{Increase in electron effective mass with carrier concentration as determined by fitting the Drude model to the La:BaSnO$_3$ IR dielectric function, independently obtained from both transmission and reflectivity spectra using vacuum FTIR spectroscopy (black circles) and spectroscopic ellipsometry (blue squares).}
		\label{fig:DOSeffmass}
	\end{figure}

	\sloppy
	Optical transmission and reflectivity spectra of La:BaSnO$_3$ films were measured by vacuum Fourier transform infrared (FTIR) spectroscopy for photon energies of \mbox{$0.1-1$~eV}~\cite{supplement}. The IR dielectric function $\epsilon(\omega)$ of La:BaSnO$_3$ is extracted by fitting theoretical transmission and reflectivity spectra to the measured ones~\cite{Heavens1960}, accounting for the dielectric function of the MgO substrate and NiO buffer layer. The IR dielectric function is described by the Drude free electron model
	\begin{equation}
	\varepsilon (\omega) = \varepsilon_{\infty} - \frac{\varepsilon_{\infty} \omega_\text{p}^2}{\omega^2 + i \gamma_\text{p} \omega}~,
	\end{equation}
	where $\varepsilon_{\infty}$ is the high frequency dielectric constant, $\omega_\text{p}$ is the plasma frequency, $\omega$ is the photon frequency and $\gamma_\text{p}$ is the broadening frequency. The parameters used in the analytical model allow for calculating the electron effective mass
	\begin{equation}
	m^*_\text{e} = \frac{n e^2}{\omega_p^2 \varepsilon_{\infty} \varepsilon_0}~,
	\end{equation}
	where $\epsilon_0$ is the vacuum dielectric constant and $n$ is the carrier concentration, which was determined from alternating current Hall measurements. The La:BaSnO$_3$ electron effective mass increases remarkably from 0.21~$m_0$ to 0.40~$m_0$ for carrier concentrations from $1.6\times10^{19}$ to \mbox{$7.3\times10^{20}$~cm$^{-3}$}, respectively~(Fig.~\ref{fig:DOSeffmass}).
	
	The energy dispersion $E$ of electrons near the conduction band minimum can be described using a first-order non-parabolicity approximation~\cite{Pisarkiewicz1989}
	\begin{equation}
	\frac{\hbar^2 k^2}{2m^*_\text{e0}} = E \left(1 + \beta E \right),
	\label{eq:Enp}
	\end{equation}
	where $\hbar$ is the reduced Planck constant, $m^*_\text{e0}$ is the electron effective mass at the conduction band minimum, $k = \left( 3 n \pi^2 \right)^{1/3} $ is the Fermi wave vector and $\beta$ is a fitting parameter describing the degree of non-parabolicity of the conduction band. After solving the quadratic equation~\eqref{eq:Enp} to obtain the electron energy $E$ and applying the relation~\cite{Hummel2011}
	\begin{equation}
	\frac{1}{m^*_\text{e}} = \frac{1}{\hbar^2 k} \frac{dE}{dk},
	\label{eq:meff}
	\end{equation}
	the effective mass is given by
	\begin{equation}
	m_\text{e}^* = m^*_\text{e0} \sqrt{1 + 2\beta~\frac{\hbar^2 k^2}{m^*_\text{e0}}}~.
	\label{eq:md}
	\end{equation}
	Fitting of equation~\eqref{eq:md} to the effective mass obtained by analysis of FTIR spectra yields $m^*_\text{e0} = 0.19~m_0$ at the conduction band minimum, in good agreement with the result from hybrid density functional theory~\cite{Scanlon2013}, and the non-parabolicity parameter $\beta = 0.56$~(eV)$^{-1}$. The increase in effective mass is remarkably larger in La:BaSnO$_3$ as compared to other transparent conducting oxides like Ga:ZnO ($\beta = 0.14$~(eV)$^{-1}$) and Sn:In$_2$O$_3$ ($\beta = 0.18$~(eV)$^{-1}$)~\cite{Fujiwara2005}, indicating a significant dependence of mobility on doping level.
	
\begin{figure}[h]
		\centering
		\includegraphics[width=12.9cm]{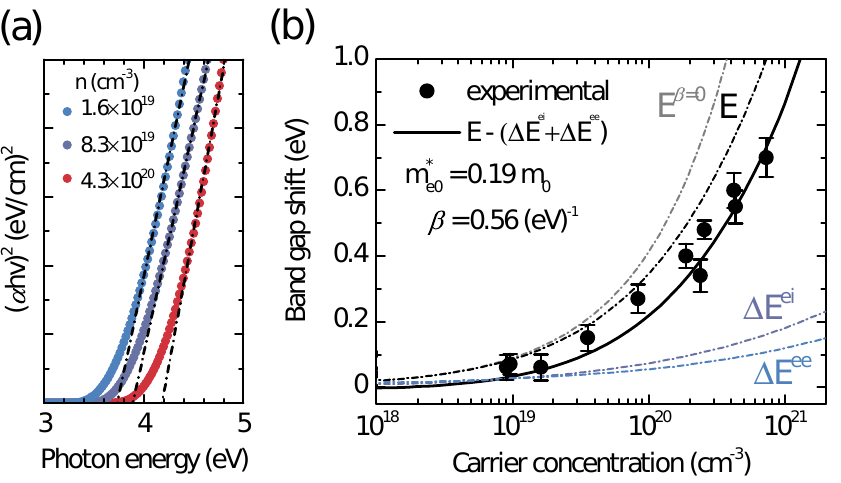}
		\caption{(a) Optical absorption spectra plotted for a direct-type transition. The optical band gap shifts from 3.71 to 4.23~eV when the carrier concentration increases from $1.6\times 10^{19}$ to $4.3\times 10^{20}$~cm$^{-3}$. (b) Increase in the La:BaSnO$_3$ optical band gap with carrier concentration due to conduction band filling effect, where $E^{\beta=0}$ is the energy of electrons assuming a non-parabolic band dispersion. The theoretical model employs the effective mass and non-parabolicity parameter derived from FTIR spectra analysis to calculate the energy dispersion $E$ of conduction band electrons and takes the band gap narrowing due to electron-ion ($\Delta E_\text{g}^\text{ei}$) and electron-electron interactions ($\Delta E_\text{g}^\text{ee}$) into account.}
		\label{fig:BMshift}
	\end{figure}

	The dependence of the La:BaSnO$_3$ effective mass on carrier concentration is further investigated by reflection spectroscopic ellipsometry using photon energies of 0.6 to 4.8~eV. First, the free carrier absorption in the IR spectral region of the dielectric function is analysed for comparison with FTIR data~\cite{supplement}. Secondly, the optical spectra were used to extract the La:BaSnO$_3$ optical band gap independently from that of the NiO buffer layer for analysis of the Burstein-Moss shift. The optical band gap was determined from a plot of $(\alpha h\nu)^2$ vs. photon energy $h\nu$, where $\alpha$ denotes the absorption coefficient, which is valid for a direct transition at the absorption edge (Fig.~\ref{fig:BMshift}(a)). Due to the conduction band filling effect, the La:BaSnO$_3$ optical absorption edge shifts from 3.71 to 4.23~eV when the carrier concentration increases from $1.6\times 10^{19}$ to $4.3\times 10^{20}$~cm$^{-3}$.
	
	With increasing carrier concentration, electron-electron and electron-impurity interactions partially compensate the Burstein-Moss shift as the conduction band edge is shifted to lower energies. The band gap narrowing due to the electron-electron interactions is given by~\cite{Berggren1981}
	\begin{equation}
	\Delta E_g^\text{ee} = \frac{e^2k}{2\pi^2 \varepsilon_\text{s} \varepsilon_0}+\frac{e^2 \lambda}{8\pi \varepsilon_\text{s} \varepsilon_0} \left(1- \frac{4}{\pi} \tan^{-1}\left(\frac{k}{\lambda} \right) \right),
	\end{equation}
	where $\varepsilon_\text{s} = 20$~\cite{Stanislavchuk2012,Niedermeier2016_arXiv_LOph} is the BaSnO$_3$ static dielectric constant, $\lambda = 2\sqrt{k/a^*_\text{B} \pi}$ is the Thomas-Fermi screening length, $a^*_\text{B} = 4\pi \varepsilon_\text{s} \varepsilon_0 \hbar^2 / m^*_\text{e} e^2$ is the effective Bohr radius and $\hbar$ is the reduced Planck constant. The band gap narrowing due to the electron-impurity interactions is given by
	\begin{equation}
	\Delta E_g^\text{ei} = \frac{ne^2}{a^*_\text{B} \varepsilon_\text{s} \varepsilon_0 \lambda^3}~.
	\end{equation}
	In total, both effects contribute to ca. 0.2~eV narrowing of the optical band gap for the highest electron concentration of $7.3\times 10^{20}$~cm$^{-3}$. After subtraction of the conduction band shifts, the Burstein-Moss shift in La:BaSnO$_3$ is described consistently with Eq.~\eqref{eq:Enp}, employing the effective mass and the non-parabolicity parameter obtained from IR spectra analysis (the solid curve in Fig.~\ref{fig:BMshift}(b)). As compared to the Burstein-Moss shift analysis for La:BaSnO$_3$ using the indirect electronic band gap determined by photoelectron spectroscopy~\cite{Lebens-Higgins2016}, the present results indicate a more pronounced non-parabolicity of the conduction band and a weaker effect of many-body electron-electron and electron-impurity interactions.
	
\begin{figure}[h]
		\centering
		\includegraphics[width=\textwidth]{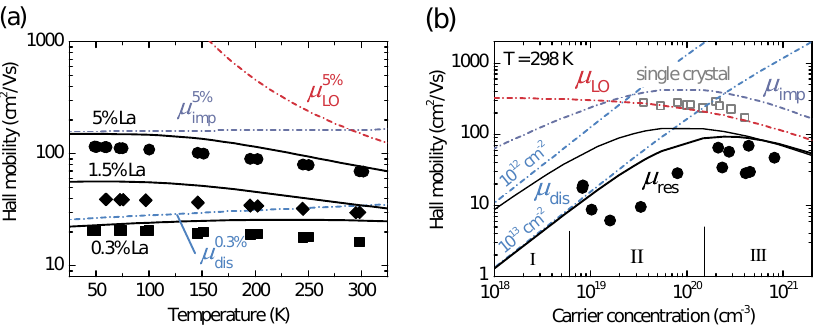}
		\caption{(a) Temperature-dependent Hall mobility of 0.3~\%, 1.5~\% and 5~\% La-doped BaSnO$_3$ films indicates degenerate, metal-like transport behaviours, which is dominated by dislocation scattering for low doping levels and ionized impurity and LO optical phonon scattering for high doping levels. Solid lines present theoretical calculation of the mobility using the three analytical scattering models according to Eqs.~\eqref{eq:u_dis}, \eqref{eq:u_ph}, \eqref{eq:u_imp} and \eqref{eq:utot}. (b) La:BaSnO$_3$ films are too resistive to measure Hall voltages for doping levels below 0.3\%~La~(I). The room temperature Hall mobility of the degenerate La:BaSnO$_3$ films is governed by dislocation scattering in the intermediate doping regime (II) and LO optical phonon and ionized impurity scattering in the high doping regime (III). The mobility of La:BaSnO$_3$ crystals is given for comparison (grey open squares, after Kim, Phys. Rev. B {\bf 86}, 165205 (2012)~\cite{Kim2012_PRB}).}
		\label{fig:mobility}
	\end{figure}

	
	After the determination of the effective mass, the mobility in La:BaSnO$_3$ can be quantitatively described using Eq.~\eqref{eq:mobility} after adopting an analytical description of the relaxation time for carrier scattering. The temperature-dependent electron transport properties for 0.3, 1.5 and and 5~at.\% La-doped films show degenerate, metal-like behaviours characterized by a constant carrier concentration~\cite{supplement} and only moderately increasing Hall mobility upon decreasing the temperature from 300 to 45~K~(Fig.~\ref{fig:mobility}(a)). Since the La (0/$+$) charge transition level in BaSnO$_3$ lies within the conduction band~\cite{Scanlon2013}, it may be assumed that all the La atoms are readily ionized to donate an electron to the reservoir of free carriers. However, for small doping levels the Hall carrier concentration is significantly reduced as compared to the La impurity concentration~\cite{supplement}. The La:BaSnO$_3$ films of less than 0.3~at.\% La impurity concentration ($n_\text{La} = 4\times 10^{19}$~cm$^{-3}$) are highly resistive, which suggests trapping of free carriers by defects in the microstructure. Above a doping level of 0.3~at.\% La, the room temperature Hall mobility increases from 18 to 70~cm$^2$/Vs at carrier concentrations from $8.3\times 10^{18}$ to $4.2\times 10^{20}$~cm$^{-3}$, respectively, but then drops at higher impurity concentrations~(Fig.~\ref{fig:mobility}(b)).
	
	The crystal mosaicity observed in the HR-XRD analysis and the microstructure in the TEM observation suggest the vertical grain boundaries as possible carrier traps. However, the activation energy of mobility, which reflects the electron transport potential barrier height and temperature-dependent scattering properties, are as low as 2.5 to 3.6~meV and significantly smaller than the thermal energy $k_\text{B}T$, where $k_\text{B}$ is the Boltzmann constant and $T$ is the absolute temperature~\cite{supplement}. Thus grain boundaries do not affect the electron transport properties at room temperature~\cite{Seto1975}.
	
	Since the BaSnO$_3$ perovskite structure consists of alternating layers of BaO and SnO$_2$, e.g. stacking faults including Ruddlesden-Popper-type ones are readily introduced into the microstructure when different crystal domains coalesce during thin film growth~\cite{Wang2015}. Such structural defects are introduced even at the exact Ba/Sn growth stoichiometry and independent from the dislocations resulting from the structural mismatch to the substrate. Therefore, dislocation scattering is investigated as the prevailing mobility-limiting transport mechanism in BaSnO$_3$ epitaxial films for carrier concentrations below $1\times 10^{20}$~cm$^{-3}$. The dislocations which may create trap states for free electrons may explain the significantly reduced doping efficiency when the La impurity concentration is comparable to or smaller than the trap density~\cite{Weimann1998}. The mobility governed by dislocation scattering in a degenerate semiconductor is described by~\cite{Look2001}
	\begin{equation}
	\mu_\text{dis} = \frac{8ea^2}{\pi hN_\text{dis}} \left( \frac{3n}{\pi} \right)^{2/3} \left(1+ \xi_0 \right)^{3/2},
	\label{eq:u_dis}
	\end{equation}
	where
	\begin{equation}
	\xi_0 = \frac{\varepsilon_\text{s} \varepsilon_0 h^2 }{m^*_\text{e} e^2 } \left( 3\pi^2 n \right)^{1/3},
	\label{eq:xi0}
	\end{equation}
	and $a$ is the BaSnO$_3$ lattice parameter, h is the Planck constant and $N_\text{dis}$ is the dislocation density. For carrier concentrations above \mbox{$1\times 10^{20}$~cm$^{-3}$}, the room temperature mobility in BaSnO$_3$ is governed by electron-phonon interactions and electron scattering at ionized La impurities~\cite{Krishnaswamy2016_arXiv}. The mobility governed by longitudinal optical (LO) phonon scattering is calculated according to~\cite{Frohlich1939,Low1953}
	\begin{equation}
	\mu_{\text{LO}_\upmu} = \frac{1}{2c_\upmu \omega_{\text{l}_\upmu}} \left(1+ \frac{c_\upmu}{6} \right)^{-2} f(c_\upmu) \left( \exp \left( \frac{\hbar \omega_{\text{l}_\upmu}}{k_\text{B}T} \right) -1 \right)
	\label{eq:u_ph}
	\end{equation}
	where $c_\upmu$ is the electron-phonon coupling constant proportional to the square root of electron effective mass $\sqrt{m_\text{e}^*}$, $\omega_{\text{l}_\upmu}$ is the frequency of the LO phonon mode and $f(c_\upmu)$ is a slowly varying function ranging from 1.0 to 1.2 for $0 < c_\upmu < 3$~\cite{Low1955}. In La:BaSnO$_3$, the three LO phonon modes with $\hbar \omega_\text{l}= 18$, 55 and 97~meV are taken into account~\cite{Stanislavchuk2012, Niedermeier2016_arXiv_LOph} to calculate an effective mobility $\mu_\text{LO}$ using the sum of reciprocals for each phonon mode, using $\mu_\text{LO}^{-1} = \sum\limits_\upmu \mu_{\text{LO}_\upmu}^{-1}$~\cite{Eagles1964}. The mobility governed by ionized impurity scattering in a degenerate semiconductor~\cite{Brooks1955,Dingle1955}, taking the non-parabolicity of the conduction band into account~\cite{Zawadzki1982}, is given by 
	\begin{equation}
	\mu_\text{imp} =\frac{3\varepsilon_\text{s}^2 \varepsilon_0^2 h^3}{m^{*2}_\text{e} e^3}~ \frac{n}{N_\text{i}}~ \frac{1}{F}~,
	\label{eq:u_imp}
	\end{equation}
	where $F$ is the screening function
	\begin{equation}
	F = \left(1+4\frac{\xi_1}{\xi_0} \left(1-\frac{\xi_1}{8} \right) \right) \ln (1+\xi_0) - \frac{\xi_0}{1+\xi_0} - 2\xi_1 \left(1-\frac{5}{16} \xi_1 \right)
	\end{equation}
	and
	\begin{equation}
	\xi_1 = 1- \frac{m^*_\text{e0}}{m^*_\text{e}}~.
	\end{equation}
	$N_\text{i}$ is the concentration of ionized impurities, which is equal to the La concentration assuming that all the La impurity atoms donate one electron each. According to the Matthiessen's rule, the resulting mobility is
	\begin{equation}
	\mu_\text{res}^{-1} = \mu_\text{dis}^{-1} + \mu_\text{LO}^{-1} + \mu_\text{imp}^{-1}~,
	\label{eq:utot}
	\end{equation}
	which describes well the dependence on temperature and carrier concentration over the entire range of investigation~(Fig.~\ref{fig:mobility}).
	
	The analysis according to equations~\eqref{eq:u_dis}, \eqref{eq:xi0}, \eqref{eq:u_ph} and~\eqref{eq:u_imp} shows that the high La:BaSnO$_3$ mobility is mainly attributed to two quantities, the small electron effective mass and the large static dielectric constant. The room-temperature effective mass increases pronouncedly from $m_\text{e}^* = 0.21$~$m_0$ to 0.40~$m_0$ for the range of the carrier concentrations investigated in this work. This is reflected by the large non-parabolicity parameter $\beta=0.56$~(eV)$^{-1}$ and noticeably reduces the mobility at the highest doping levels concurrent with an increase in the polaron mass. However, the large dielectric constant $\varepsilon_\text{s} = 20$~\cite{Stanislavchuk2012,Niedermeier2016_arXiv_LOph} of BaSnO$_3$ promotes screening of the Coulomb potential of charged dislocations and ionized La impurities~\cite{Kim2012_PRB}. The square dependence of mobility governed by impurity scattering on dielectric constant shows that in contrast to other high mobility wide band gap semiconductors like ZnO ($\varepsilon_\text{s}^\perp = 7.4$~\cite{Yoshikawa1997}), In$_2$O$_3$~($\varepsilon_\text{s}=8.9$~\cite{Hamberg1986}) and GaN ($\varepsilon_\text{s}^\perp = 9.5$~\cite{Barker1973}), BaSnO$_3$ may exhibit a large mobility even at unusually high carrier concentrations because impurity scattering does not become significant. Furthermore, electron-phonon scattering is less pronounced as compared to other polar oxides such as SrTiO$_3$~($\sim$10~cm$^2$/Vs~\cite{Verma2014}) and Ga$_2$O$_3$ ($\sim$115~cm$^2$/Vs~\cite{Ghosh2016}), allowing to achieve an extraordinary high $\sim$300~cm$^2$/Vs mobility even at room temperature~\cite{Krishnaswamy2016_arXiv, Niedermeier2016_arXiv_LOph}.
	
	Further progress in the development of BaSnO$_3$ epitaxial films for application in transparent oxide electronics may be realized not only by considering the lattice mismatch with the substrate, but also by optimising epitaxial growth techniques to reduce dislocation densities and domain boundaries which are inherent to the perovskite structure. Selective-area growth and epitaxial lateral overgrowth techniques as successfully applied in III-nitride semiconductor technology~\cite{Hiramatsu2001} may present effective methods for achieving higher mobilities by reducing the dislocation density of the present BaSnO$_3$ epitaxial films.
	
	
	In conclusion, degenerate perovskite BaSnO$_3$ exhibits an extraordinary high room temperature mobility attributed to an electron effective mass as small as 0.19~$m_0$. An in-depth investigation of the infrared free carrier absorption and the Burstein-Moss shift yields a significant dependence of electron effective mass on doping level, concurrent with a pronounced non-parabolicity of the conduction band. The electron effective mass dependence on doping level is employed to quantitatively describe the scattering mechanisms in degenerate BaSnO$_3$ films over the entire range of doping levels and as a function of temperature. The current room temperature mobility limits in epitaxial films are determined by scattering at dislocations at low doping levels, and ionized impurity scattering and electron-phonon interactions at high doping levels. The large dielectric constant of BaSnO$_3$ facilitates the screening of charged defects and ionized impurities more than in other transparent semiconductor and electron-phonon scattering is less pronounced as compared to other polar oxide semiconductors, resulting in an enhanced room-temperature mobility even at unusually high carrier concentrations.
	
	\section*{Acknowledgements}
	
	We thank Dr J. Jia and Prof Y. Shigesato at Aoyama Gakuin University for assistance with spectroscopic ellipsometry measurements. C. A. Niedermeier, S. Rhode and M. A. Moram acknowledge support from the Leverhulme Trust via M. A. Moram's Research Leadership Award (RL-0072012).  M. A. Moram acknowledges further support from the Royal Society through a University Research Fellowship. The work at Tokyo Institute of Technology was supported by the Ministry of Education, Culture, Sports, Science and Technology (MEXT) Element Strategy Initiative to Form Core Research Center. H. Hiramatsu was supported by the Japan Society for the Promotion of Science (JSPS) through a Grant-in-Aid for Scientific Research on Innovative Areas "Nano Informatics" Grant Number 25106007, and Support for Tokyotech Advanced Research (STAR).

\newpage


\renewcommand{\thetable}{S\Roman{table}} 
\renewcommand{\thefigure}{S\arabic{figure}}%
\renewcommand{\thesection}{\Roman{section}}
\renewcommand{\thesubsection}{\Alph{subsection}}
\renewcommand{\thefootnote}{\fnsymbol{footnote}}

\newcolumntype{C}[1]{>{\centering\arraybackslash}m{#1}}
\newcolumntype{L}[1]{>{\raggedright\let\newline\\\arraybackslash\hspace{0pt}}m{#1}}
\linespread{1.7} 
\setlength{\parindent}{10pt}

	\thispagestyle{empty} 
	\begin{center} \linespread{1.2} \LARGE \sfb Supplementary Material: \linebreak {\LARGE \sfb Electron effective mass and mobility limits in degenerate perovskite stannate BaSnO$_3$}\\
	\end{center}

	\vspace{0.5cm}
	
	\begin{flushleft}
		
		\begingroup\linespread{1.0}\selectfont
		Christian A. Niedermeier$^{1,2,3}\footnote{Corresponding author, e-mail: c.niedermeier13@imperial.ac.uk}$, Sneha Rhode$^{1}$, Keisuke Ide$^{2}$, Hidenori Hiramatsu$^{2,3}$, Hideo Hosono$^{2,3}$, Toshio Kamiya$^{2,3}$, Michelle A. Moram$^{1}$
		
		$^1$Department of Materials, Imperial College London, Exhibition Road, London, SW7~2AZ, UK
		
		$^2$Laboratory for Materials and Structures, Tokyo Institute of Technology, Mailbox R3-4, 4259 Nagatsuta, Midori-ku, Yokohama, 226-8503, Japan
		
		$^3$Materials Research Center for Element Strategy, Tokyo Institute of Technology, 4259 Nagatsuta, Midori-ku, Yokohama, 226-8503, Japan
		
		\endgroup
		
	\end{flushleft}
	
	\setcounter{figure}{0}
	
\section{Thin Film Growth and Structural Characterization}

Polycrystalline La$_\text{x}$Ba$_\text{1-x}$SnO$_3$ (x = 0, 0.01, 0.03 and 0.07) ceramic targets for pulsed laser deposition (PLD) were prepared by calcination of stoichiometric mixtures of high purity La$_2$O$_3$, BaCO$_3$ and SnO$_2$ powders at $1200\,^{\circ}\mathrm{C}$ for 12~h, reground and sintered at $1450\,^{\circ}\mathrm{C}$ for $48$~h. Multi-phase Rietveld refinement of the X-ray diffraction (XRD) pattern of the 7~at.\% La:BaSnO$_3$ target showed that it is composed of two phases, 96.9~wt.\% La$_\text{0.04}$Ba$_\text{0.96}$SnO$_3$ (space group Pm-3m, ICSD 100792) and 3.1~wt.\% La$_2$Sn$_2$O$_7$ (spacegroup Fd-3m, ICSD 167144), due to the limited La solubility in BaSnO$_3$~\cite{Yasukawa2010}. A NiO target was prepared by sintering high purity NiO powder at $1200\,^{\circ}\mathrm{C}$ for 5~h.

200-nm La:BaSnO$_3$ thin films were deposited on a 50-nm NiO buffer layer on MgO(100) single crystal substrates by pulsed laser ablation of the targets using a KrF excimer laser (248~nm) at a pulse frequency of 8~Hz. The NiO buffer layer was deposited at a substrate temperature of $700\,^{\circ}\mathrm{C}$, O$_2$ pressure of 0.7~Pa and laser beam fluence of 2.3~J/cm$^2$. The La:BaSnO$_3$ thin film was deposited at a substrate temperature of $800\,^{\circ}\mathrm{C}$, O$_2$ pressure of 20~Pa and laser beam fluence of 1.8~J/cm$^2$. The La doping concentration in the BaSnO$_3$ films was varied from $0-7$~at.\%~in the Ba site by co-ablation of two La$_\text{x}$Ba$_\text{1-x}$SnO$_3$ ceramic targets of different compositions $x$ using a multi-target carousel.\\

\begin{figure}[hb]
	\centering
	\sf{\includegraphics[height=6.5cm]{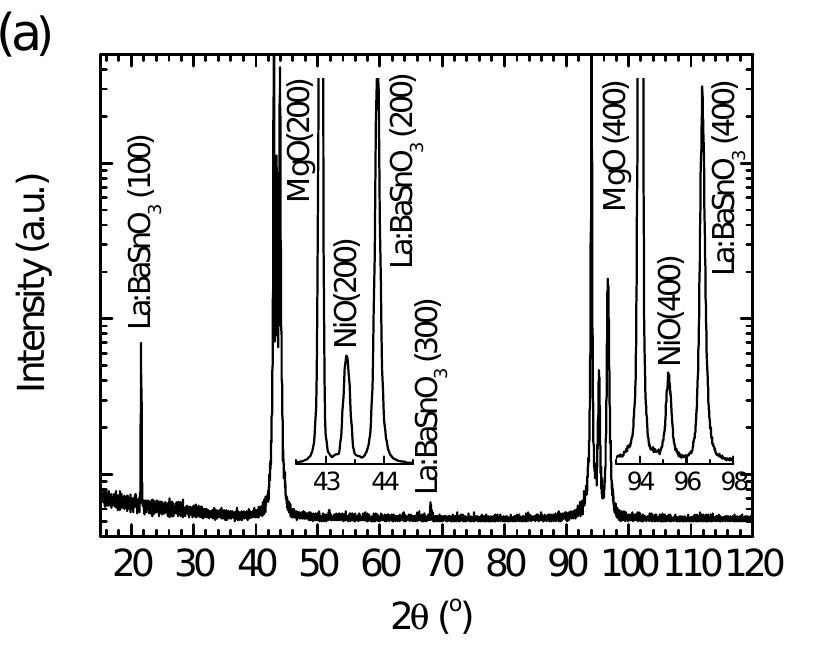}} \qquad 
	\sf{\includegraphics[height=6.5cm]{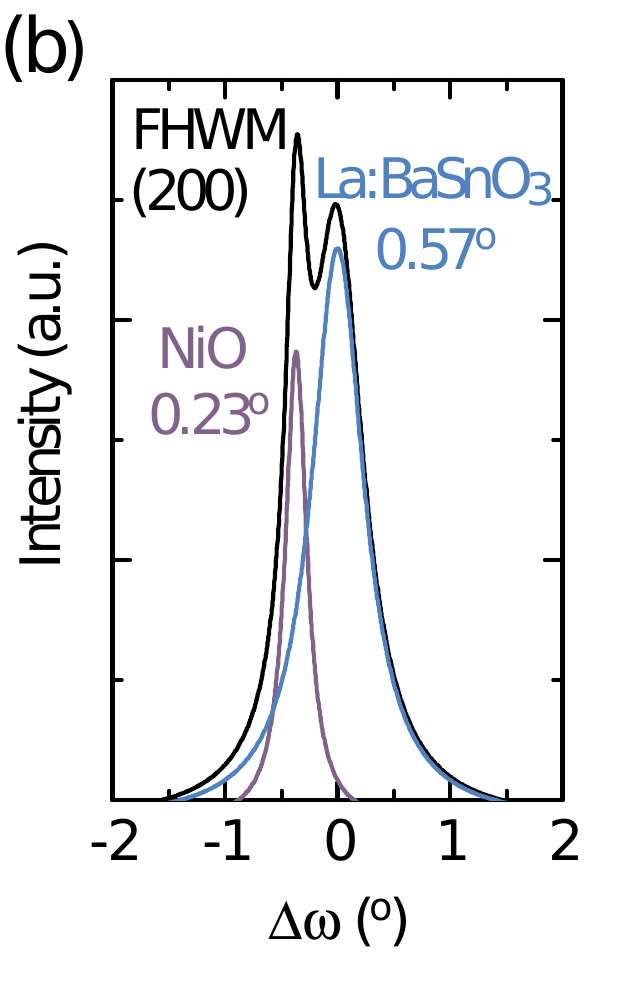}} \qquad 
	\caption{(a) Out-of-plane HR-XRD pattern of La:BaSnO$_3$ thin film on NiO-buffered MgO substrate in a logarithmic plot, indicating the single-phase and 100 oriented epitaxial growth. Insets show magnified views of the 200 and 400 diffraction peaks on a linear scale. (b) The 0.57$^{\circ}$ FWHM rocking curve of the La:BaSnO$_3$ 200 diffraction demonstrates epitaxial growth with a moderate degree of mosaicity.}
	\label{fig:xrd}
\end{figure}

The crystal structures of the films were investigated by high resolution X-ray diffraction (HR-XRD) employing a Philips MRD diffractometer with a monochromatic Cu K$_{\alpha1}$ (1.5406~\AA) X-ray source and parallel beam optics. The out-of-plane diffraction pattern of the La:BaSnO$_3$/NiO heterostructure demonstrates single-phase epitaxy along the [100] crystallographic direction for both layers~(Fig.~\ref{fig:xrd}(a)). By applying a NiO buffer layer ($\sim$4.18~\AA), the lattice mismatch with the La:BaSnO$_3$ thin film ($\sim$4.12~\AA) is reduced to less than 1.4~\%, facilitating the epitaxy on MgO substrates. The 0.57$^{\circ}$ full width at half maximum (FWHM) of the out-of-plane 200 La:BaSnO$_3$ diffraction peak obtained after deconvolution from the corresponding NiO diffraction peak indicates a moderate degree of mosaicity (Fig.~\ref{fig:xrd}(b)).

\newpage
The cross-sectional electron transmission microscopy (TEM) specimen was prepared by grinding, polishing and dimpling until the specimen thickness was below 10~$\upmu$m, followed by Ar ion milling using a PIPS Ion miller (Gatan USA). Conventional and high resolution TEM was performed using a JEOL 2100 microscope equipped with a field emission gun operating at 200~keV.

\section{Optical Characterization}

\begin{figure}[hp]
	\centering
	\includegraphics[width=8cm]{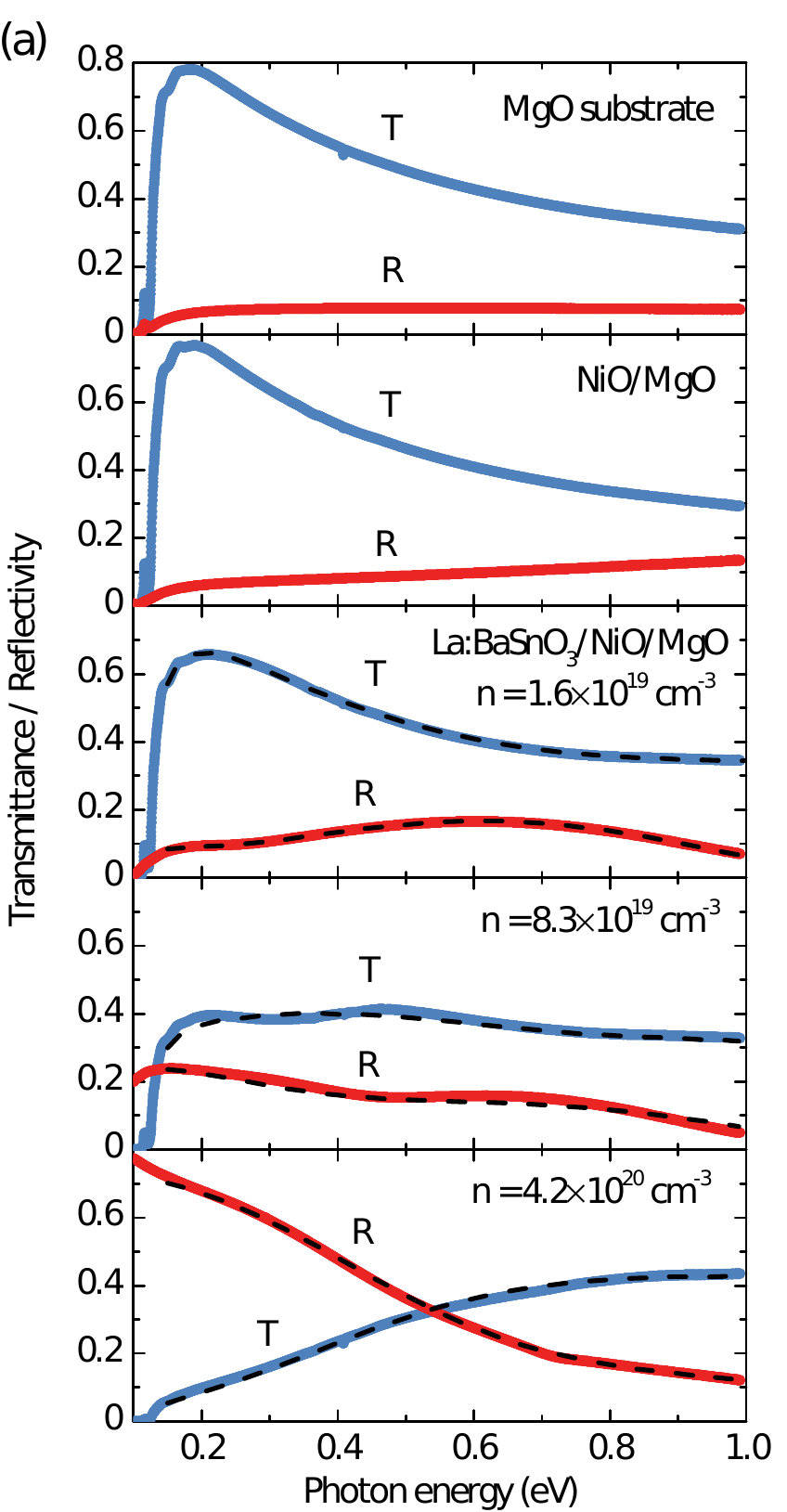}
	\includegraphics[width=8cm]{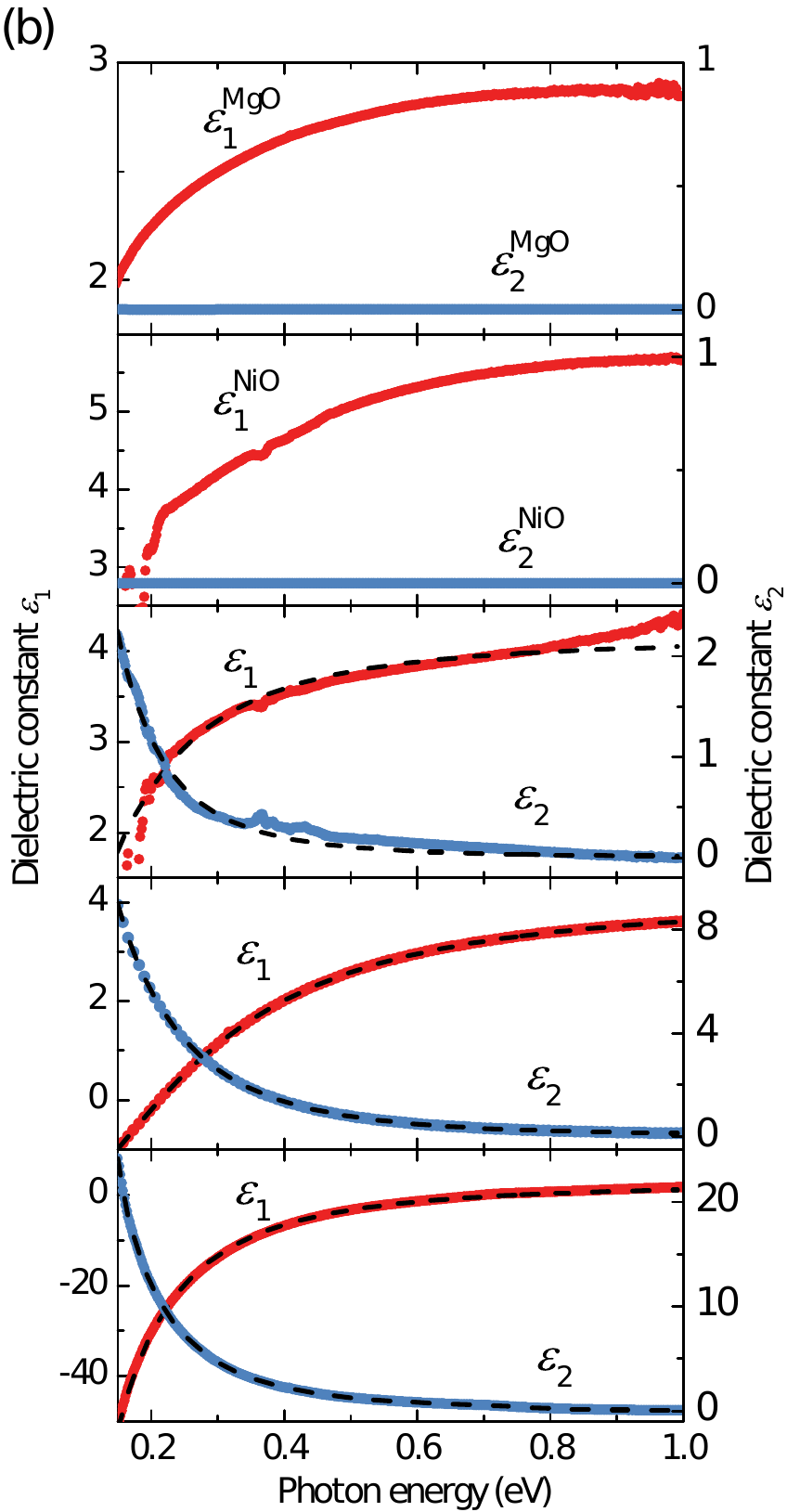}
	\caption{(a) Optical IR transmittance (blue) and reflectivity (red) spectra of bare MgO substrate, NiO film and La:BaSnO$_3$/NiO bilayer structure on MgO substrate. Carrier concentrations in La:BaSnO$_3$ were $1.6 \times 10^{19}$~cm$^{-3}$, \mbox{$8.3\times10^{19}$~cm$^{-3}$} and $4.2 \times 10^{20}$~cm$^{-3}$ as determined by Hall effect measurements. (b)~Corresponding dielectric functions of MgO, NiO and La:BaSnO$_3$ obtained by a point-by-point analysis of the IR spectra. Dashed lines to the La:BaSnO$_3$ optical spectra indicate theoretical fitting curves employing the Drude model for free carrier absorption.}
	\label{fig:ftir}
\end{figure}

The optical transmittance and reflectivity of the La:BaSnO$_3$ thin films were recorded with a Bruker Vertex 70v vacuum Fourier transform infrared (FTIR) spectrometer for photon energies of $0.1-1$~eV. By comparison with the spectrum of a 50-nm NiO film on a MgO substrate, the free carrier absorption in La:BaSnO$_3$ below 0.4~eV energy is observed even for a small carrier concentration of $1.6 \times 10^{19}$~cm$^{-3}$ (Fig.~\ref{fig:ftir}(a)). When the doping level is increased to $4.2 \times 10^{20}$~cm$^{-3}$, the La:BaSnO$_3$ film shows a significantly enhanced IR absorption and reflectivity. The corresponding IR dielectric function of La:BaSnO$_3$ for all carrier concentrations is calculated by a point-by-point calculation of the dielectric function and fitting theoretical transmittance and reflectivity spectra to the measured ones~\cite{Heavens1960} with accounting for the dielectric function of the MgO substrate and NiO buffer layer~(Fig.~\ref{fig:ftir}(b)). The Drude free electron model is fitted to the dielectric function for determination of the plasma frequency $\omega_\text{p}$ and electron effective mass~$m_\text{e}^*$~(Table~\ref{tab:ftir}).

\begin{table}[h]
	\centering
	\caption{Electron effective mass $m_\text{e}^*$ for La:BaSnO$_3$ films investigated by FTIR spectroscopy presented in Fig.~\ref{fig:ftir} with different carrier concentrations $n$. The fitting parameters of the Drude free electron model to the IR dielectric function are the high frequency dielectric constant $\varepsilon_\infty$, the plasma frequency $\omega_\text{p}$ and the broadening frequency $\gamma_\text{p}$.}
	\begin{tabular}{c*{4}{C{2cm}}} \hline \hline
		$n$ & $\varepsilon_\infty$ & $\hbar \omega_\text{p}$& $\hbar \gamma_\text{p}$& $m_\text{e}^*$\\
		(cm$^{-3}$) & - & (eV) & (eV) &(m$_0$)\\ \hline
		$1.6 \times 10^{19}$ & 3.7 & 0.17 & 0.12 & 0.21 \\
		$8.3 \times 10^{19}$ & 3.8 & 0.35 & 0.24 & 0.25 \\
		$4.2 \times 10^{20}$ & 3.2 & 0.73 & 0.10 & 0.35 \\ \hline \hline
	\end{tabular}
	\label{tab:ftir}
\end{table}

\begin{figure}[hp]
	\centering
	\includegraphics[height=12cm]{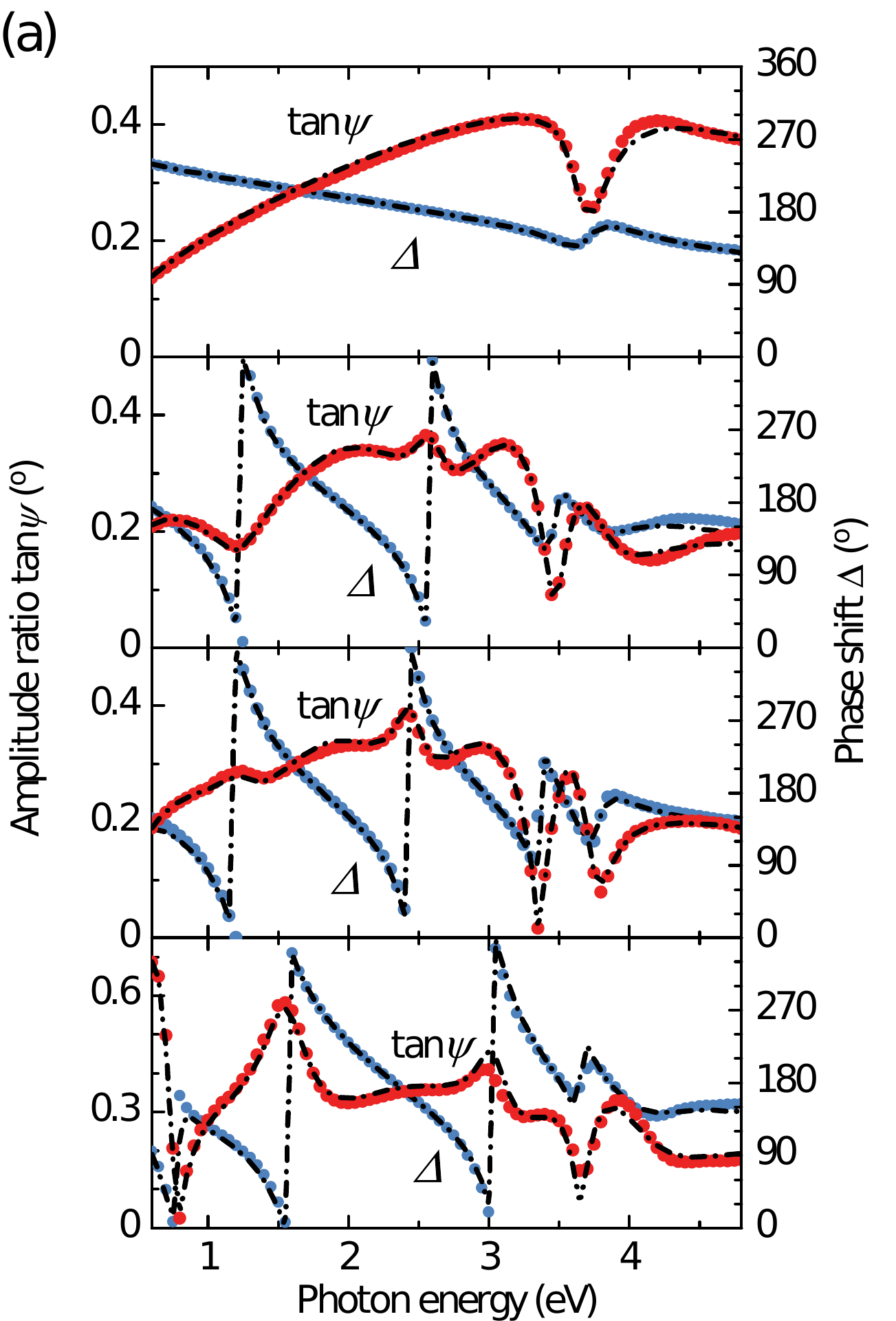}
	\includegraphics[height=12cm]{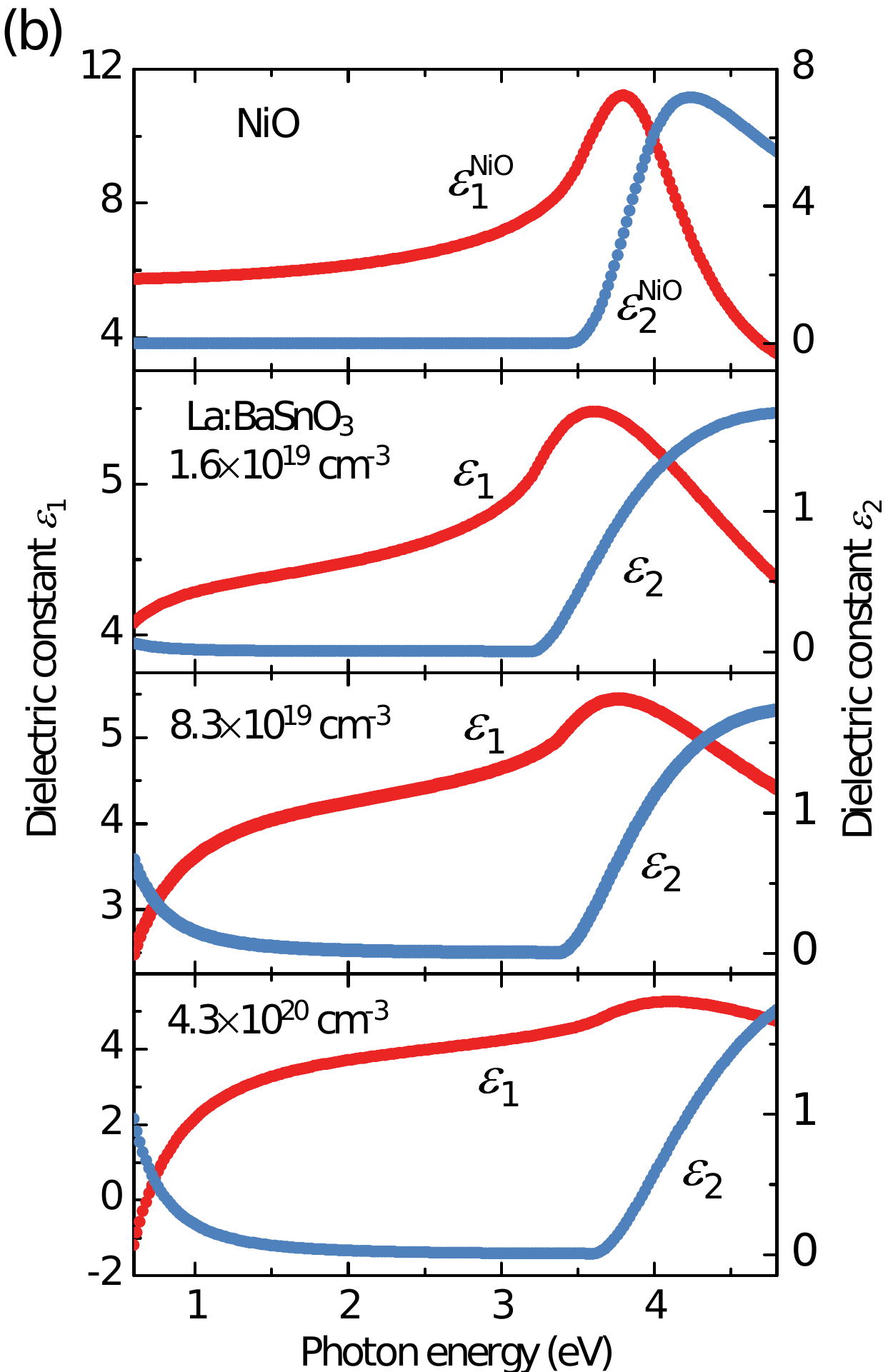}
	\caption{(a) Ellipsometry spectra of amplitude ratio $\tan\Psi$ (red) and phase shift $\Delta$ (blue) of NiO film on MgO substrate and La:BaSnO$_3$/NiO bilayer structure on MgO substrate. Carrier concentrations in La:BaSnO$_3$ were $1.6 \times 10^{19}$~cm$^{-3}$, \mbox{$8.3\times10^{19}$~cm$^{-3}$} and \mbox{$4.3 \times 10^{20}$~cm$^{-3}$} as determined by Hall effect measurements. (b)~Corresponding dielectric functions of NiO and La:BaSnO$_3$ obtained by fitting the Tauc-Lorentz dispersion function and the Drude model for free carrier absorption to the ellipsometry spectra of amplitude ratio $\tan \Psi$, phase shift $\Delta$ (dashed lines in (a)) and the pseudo-dielectric function.}
	\label{fig:ellipsometry}
\end{figure}

Ellipsometry spectra were recorded using a Jobin Yvon UVISEL spectrophotometer equipped with a phase modulator, at 60$^\circ$ angle of incidence and in the spectral range from $0.6-4.8$ eV~(Fig.~\ref{fig:ellipsometry}(a)). The Tauc-Lorentz dispersion function is employed to describe the dielectric function of the NiO buffer layer~\cite{Jellison1996}
\begin{equation}
\varepsilon_2(E) = 
\begin{cases}
\frac{A E_0 C (E-E_\text{g})^2}{(E^2-E_0^2)^2 + C^2 E^2} \frac{1}{E},& E > E_\text{g}\\
0,              & E \leq E_\text{g} \\
\end{cases}
\end{equation}
and
\begin{equation}
\varepsilon_1(E) = \varepsilon_\infty + \frac{2}{\pi} P\int_{E_\text{g}}^\infty \frac{\xi \epsilon_2(\xi)}{\xi^2-E^2} d\xi ,
\end{equation}
where $\varepsilon_\infty$ is the high frequency dielectric constant, $E_\text{g}$ is correlated to the band gap energy, $A$ is a fitting parameter in units of energy, $E_0$ is the transition energy for the Lorentz contribution, $C$ is the broadening energy and $P$ denotes the Cauchy principal part of the integral. The dielectric function of La:BaSnO$_3$ films was described using a combination of the Tauc-Lorentz dispersion function and Drude model~(Table~\ref{tab:ellipsometry}). With increasing carrier concentration from $1.6 \times 10^{19}$ to $4.3 \times 10^{20}$~cm$^{-3}$, the dielectric function indicates an increase in free carrier absorption below 1~eV and a shift of the optical transition near 3.7~eV to higher energies~(Fig.~\ref{fig:ellipsometry}(b)). To determine the La:BaSnO$_3$ optical band gap, the absorption coefficient was calculated from the dielectric function using
\begin{equation}
\alpha = \frac{4\pi k_\text{a}}{\lambda_\text{ph}}~,
\end{equation}
where $k_\text{a}$ is the optical absorption constant and $\lambda_\text{ph}$ is the photon wavelength.

\begin{table}[h]
	\centering
	\caption{Theoretical parameters used to model the dielectric functions of NiO and La:BaSnO$_3$ using the Tauc-Lorentz dispersion function, where $\varepsilon_\infty$ is the high frequency dielectric constant, $E_\text{g}$ is correlated to the band gap energy, $A$ is a fitting parameter in units of energy, $E_0$ is the transition energy for the Lorentz contribution,  and $C$ is the broadening energy. The Drude model for free carrier absorption contains the plasma frequency $\omega_\text{p}$ and broadening frequency $\gamma_\text{p}$, and $m_\text{e}^*$ indicates the electron effective mass.}
	\begin{tabular}{cc*{8}{C{1.02cm}}} \hline \hline
		material & $n$ & $\varepsilon_\infty$ & $E_\text{g}$ & $A$ & $E_0$ & $C$ & $\hbar \omega_\text{p}$ & $\hbar \gamma_\text{p}$ & $m_\text{e}^*$ \\
		& (cm$^{-3}$) & - & (eV) & (eV) & (eV) & (eV) & (eV) & (eV) & ($m_0$)\\ \hline
		NiO & & 3.4 & 3.5 & 390 & 3.8 & 0.92 &  &  &  \\
		La:BaSnO$_3$ & $1.6 \times 10^{19}$ & 3.4 & 3.2 & 310 & 2.4 & 1.5 & 0.17$^\text{a}$ & 0.12$^\text{a}$ & 0.21$^\text{a}$ \\ 
		La:BaSnO$_3$ & $8.3 \times 10^{19}$ & 3.2 & 3.4 & 340 & 2.7 & 1.4 & 0.39 & 0.19 & 0.23 \\
		La:BaSnO$_3$ & $4.3 \times 10^{20}$ & 2.9 & 3.6 & 530 & 2.4 & 1.6 & 0.81 & 0.11 & 0.31 \\ \hline \hline
		\multicolumn{10}{L{15.7cm}}{\footnotesize{$^\text{a}$These values were adopted from FTIR analysis since free carrier absorption above 0.6~eV is weak.}} \\
	\end{tabular}
	\label{tab:ellipsometry}
\end{table}

\newpage

\section{Electron Transport Properties}

Alternating current Hall measurements of La:BaSnO$_3$ films at room temperature were performed with a LakeShore 8400 system (Toyo Co.) using the four-terminal method with sputtered Au/Ti contacts in van der Pauw geometry. The Hall carrier concentration was accurately determined using the film thickness derived from spectroscopic ellipsometry analysis. Temperature-dependent electrical transport properties were measured with a ResiTest 8300 system (Toyo Co.) in the range from 300 to 45~K in He atmosphere.

\begin{figure}[tp]
	\centering
	\includegraphics[width=12.9cm]{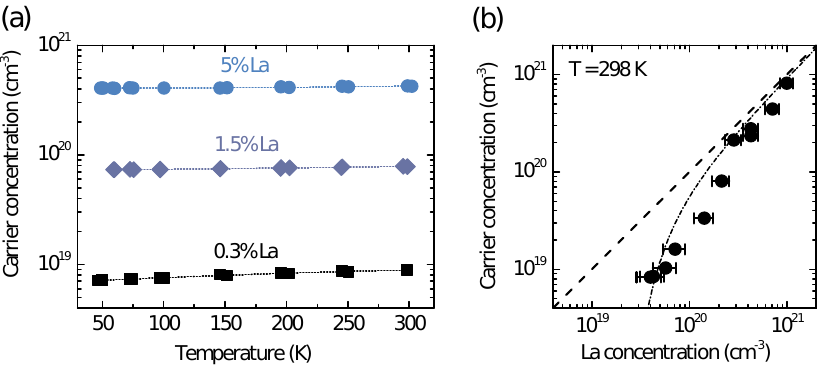}
	\caption{(a) The carrier concentration of 0.3, 1.5 and 5~at.\% La:BaSnO$_3$ films remains constant with varying temperature and indicates degenerate, metal-like transport behaviours. (b) Comparison of the measured carrier concentration obtained at room temperature and actual La impurity concentration. The straight dashed line indicates 100\% doping efficiency.}
	\label{fig:laconc}
\end{figure}

\sloppy
The carrier concentration of 0.3, 1.5 and 5~at.\% La:BaSnO$_3$ films remains constant with varying temperature and indicates degenerate, metal-like transport behaviours~(Fig.~\ref{fig:laconc}(a)). For small doping levels, the Hall carrier concentration is significantly reduced as compared to the actual La impurity concentration in the La:BaSnO$_3$ films, which suggests trapping of free carriers by defects in the microstructure~(Fig.~\ref{fig:laconc}(b)).

\begin{figure}[h]
	\centering
	\includegraphics[width=8.5cm]{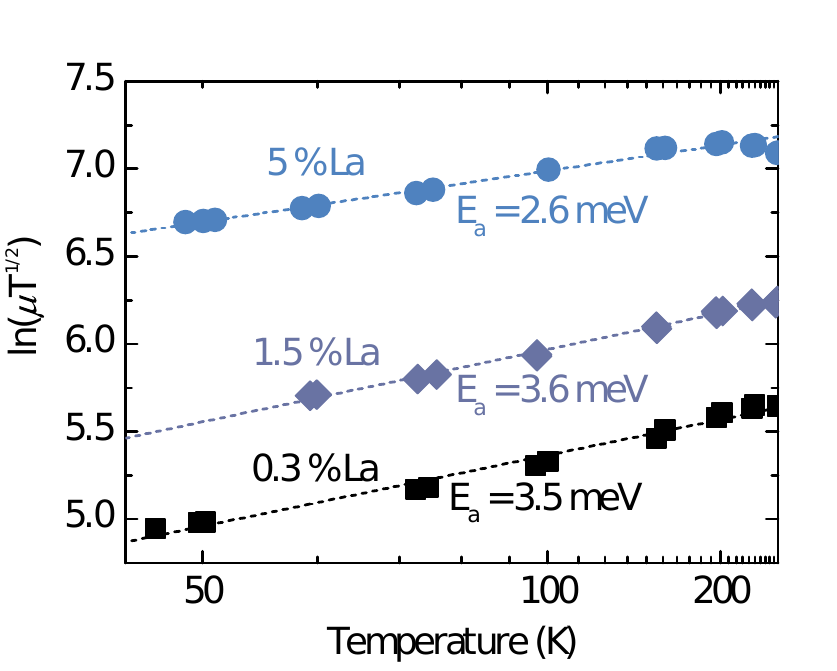}
	\caption[Arrhenius plot for evaluation of the activation energy of electron mobility in La:BaSnO$_3$ films.]{Arrhenius-type plot of $\ln (\mu T^\frac{1}{2})$ for evaluation of the activation energy of electron mobility in La:BaSnO$_3$ films.}
	\label{fig:gb_scattering}
\end{figure}

The temperature-dependent mobility for La:BaSnO$_3$ films was evaluated by employing an Arrhenius-type equation to determine the activation energy of electron mobility according to the Seto model~\cite{Seto1975} described by
\begin{equation}
\mu_\text{gb} = \text{const.} \left( \frac{1}{k_\text{B}T} \right)^{1/2} \exp \left(- \frac{E_\text{a}}{k_\text{B}T} \right),
\end{equation}
where $k_\text{B}$ is the Boltzmann constant and $T$ is the absolute temperature. The activation energy of $E_\text{a} = 2.5$ to 3.6~meV is significantly smaller than the thermal energy $k_\text{B}T$, and thus grain boundaries do not affect the electron transport properties at room temperature (Fig.~\ref{fig:gb_scattering}).

	\newpage
	

\end{document}